\documentclass[twocolumn,preprintnumbers,amssymb,pra]{revtex4-1}
\usepackage{pgfplots}
\usepackage{latexsym,epsfig}
\usepackage{graphicx}
\usepackage{dcolumn}
\usepackage[normalem]{ulem}
\usepackage{amsthm,amsmath}

\begin{document}
\title{Investigating properties of Cl$^-$ and Au$^-$ ions using relativistic many-body methods}

\author{B. K. Sahoo}

\email{bijaya@prl.res.in}

\affiliation{Atomic, Molecular and Optical Physics Division, Physical Research Laboratory, Navrangpura, Ahmedabad-380009, India}

\date{Received date; Accepted date}

\begin{abstract}
We investigate ground state properties of singly charged chlorine (Cl$^-$) and gold (Au$^-$) negative ions by employing four-component relativistic 
many-body methods. In our approach, we attach an electron to the respective outer orbitals of chlorine (Cl) and gold (Au) atoms to determine
the Dirac-Fock (DF) wave functions of the ground state configurations of Cl$^-$ and Au$^-$, respectively. As a result, all the single-particle 
orbitals see the correlation effects due to the appended electron of the negative ion. After obtaining the DF wave functions, lower-order 
many-body perturbation methods, random-phase approximation, and coupled-cluster (CC) theory in the singles and doubles approximation are applied to 
obtain the ground state wave functions of both Cl$^-$ and Au$^-$ ions. Then, we adopt two different approaches to the CC theory -- a perturbative 
approach due to the dipole operator to determine electric dipole polarizability and an electron detachment approach in the Fock-space framework to
estimate ionization potential. Our calculations are compared with the available experimental and other theoretical results.
\end{abstract}

\maketitle

\section{Introduction}

A number of stable negative atomic ions have been observed in the laboratories \cite{dudinikov,sasao}. Their spectroscopic and scattering properties
are of immense interest to both the experimentalists and theoreticians \cite{champeau1,champeau2}. It is well known fact that the Sun looks yellow
due to the black-body radiation from H$^-$ at the temperature $T=5780$ K \cite{wildt}. Another prominent example is, radiation from night-sky is 
observed due to the reaction of O$^-$ with O$_2^+$ and $N_2^+$ ions \cite{hewson}. Generally Penning traps are used to store the negative ions in the 
laboratory \cite{heinicke,kwon}, but Paul traps combined with time-resolved detection techniques are also useful to investigate photo-detachment 
processes of electrons \cite{champeau1,champeau2}. The Penning trap provides large magnetic field to analyze its effect on the negative ions, while 
the Paul trap offers better signal-to-noise ratio to perform high-precision measurements of spectroscopic properties. Though conducting 
experiments with negative ions are precarious relative to positively charged ions, there are still a number of negative ions undertaken in the 
laboratories for investigations. Some of the prominent negative atomic ions that are experimentally probed include H$^-$, Li$^-$, B$^-$, C$^-$, 
Al$^-$, Ca$^-$, Cu$^-$, Si$^-$, Cl$^-$, Au$^-$ and others (please see reviews in Refs. \cite{champeau1,champeau2,watanabe,feigerle}). Due to a lot of
demand, negative atomic ion physics are being reviewed from time to time since 1970s. Massey was one of the first persons to update the information 
about the negative ions in a monograph \cite{massey1} followed by a review article \cite{massey2}. The progresses made in the negative ion physics 
during 1980s were discussed by Bates \cite{bates}, Esaulov \cite{esaulov}, Schulz \cite{schulz} among many others. The latest review article by 
Andersen covers a wide range of topics relevant to the negative ion physics \cite{andersen}.

Owing to the complication in the experimental set up, only a few selective spectroscopic properties of the negative ions have been measured among 
which electron affinity (EA) or negative of the ionization potential (IP) of the outermost electron is the most common \cite{massey1,massey2}. A 
large number of studies are focused on the photo-detachment cross-sections using various techniques \cite{champeau1,champeau2,watanabe,ivanov,
stapelfeldt}. The typical energy levels of negative ions are quite different than their isoelectronic neutral atoms. There is only little knowledge 
revealed about the energetically excited states of negative atomic ions, but the general perception is that these states lie just above the ground 
state of the parent neutral atom. Recent studies reveal that some of the negative ions such as lanthanide sequence possess bound excited states 
\cite{Cerchiari}. Thus, these states are anticipated to be extremely short-lived. Unlike the Coulomb interactions that are solely 
responsible for binding electrons in neutral atoms and positively charged ions, the excess electron(s) in negative ions are known to be bound by 
short-range potentials \cite{andersen}. As a result, negative ions exhibit many exotic properties that are totally different from neutral atoms
and positively charged ions.

Theoretical studies of spectroscopy properties of atomic negative ions are very interesting to test the validity of quantum many-body methods.
The extrapolated EA values from the IPs of neutral atoms and the positive ions suggests that negative ions for the elements like He, N, Ne, Mg, or Ar 
ions cannot exist, but this has been disproved later \cite{andersen}. Therefore, it is imperative to apply potential quantum many-body methods from 
the first principle to study the properties of negative ions. A number of methods such as many-body perturbation theory (MBPT), multi-configuration 
Hartree-Fock (MCHF) method, random-phase approximation (RPA), R-matrix approach including coupled-cluster (CC) theory have been employed to 
investigate atomic properties and scattering cross-sections of negative ions \cite{champeau1,champeau2,wang}. It is still challenging to match the 
theoretical values with the experimental results using many-body calculations even for the basic property like EA. Most of the previous calculations 
are carried out in the non-relativistic theory framework and some cases the relativistic effects are estimated approximately \cite{neogrady}. 
There are also relativistic calculations in the negative ions reported in Refs. \cite{dzuba1,dzuba2,Verbockhaven}. The CC 
theory is considered to be the gold standard of electronic structure calculations in many-electron systems \cite{cizek,bartlett}. It captures electron
correlation effects to a much better extent than other many-body methods at the given level of approximation. Therefore, consideration of 
relativistic CC (RCC) methods are the natural choices to investigate both the relativistic and electron correlation effects in the determination of 
properties of atomic systems in general and of negative ions in particular.

Accurate evaluation of electric dipole polarizabilities ($\alpha_d$) of negative ions have been paid less attention. Their applications in crystals 
are tremendous as they help to find out mobility of negative ions due to external electric fields. The sizes of crystals can be estimated with the 
knowledge of $\alpha_d$ values of their negative ions \cite{bisrya,solomonik,volykhin,mahan}. $\alpha_d$ values of ions are used as key parameters in 
the explanation of the Hofmeister series -- the systematic trend of different ions with the same valency in their ability to precipitate in 
macromolecules from aqueous solutions \cite{zhuang} and they can be useful to analyze the behaviors of negative ions in external static fields, 
which are manifested in the threshold photo-detachment studies \cite{gibson}. Measurements of $\alpha_d$ values of negative ions are extremely difficult 
due to which only a very limited number of theoretical studies on these quantities are carried out thus far. Many of these calculations are available only for a few electron
negative ions \cite{kar,jiao,bhatia}. Theoretical studies on $\alpha_d$ values of a number of heavier negative ions are reported by Sadlej and coworkers 
\cite{diercksen,kello} by employing a variety of methods including the CC methods. They have also highlighted unusually large contributions from the 
relativistic effects to the determination of $\alpha_d$ values. But the relativistic effects were estimated by taking the spin-averaged Douglas-Kroll 
atomic Hamiltonian with no-pair (DKnp) approximation.

In this work, we intend to investigate the $\alpha_d$ values and IPs of Cl$^-$ and Au$^-$ negative ions by considering the four-component 
Dirac-Coulomb (DC) Hamiltonian at different levels of approximations in the many-body methods. We have deliberately selected these two candidates to analyze the electron correlation trends in the above properties.
Cl$^-$ is isoelectronic to Ar noble gas atom, whereas Au$^-$ is isoelectronic to Hg atom. Our previous calculations of $\alpha_d$ values in Ar 
\cite{yashpal1} and Hg \cite{yashpal2,bijaya1} atoms show a very contrast correlation contribution trends at different levels of approximations in
the many-body methods. Since the outermost electron in a negative ion is very weakly bound, the electron correlation effects can behave completely
different way in Cl$^-$ than Ar, and so as between Au$^-$ and Hg. This can be demonstrated by evaluating $\alpha_d$ values of Cl$^-$ and Au$^-$ by applying
methods similar to that were employed earlier to determine $\alpha_d$ values of Ar and Hg atoms, and making comparative analyses. There are precise 
measurements of EA of Cl$^-$ reported in Refs. \cite{trainham,hanstorp}. An interesting study together with experimental and theoretical methods was carried
out to infer mass shift from the EAs of negative ions of chlorine isotopes \cite{berzinsh}, following which another theoretical work was devoted to 
explain the discrepancy between the previous experimental and theoretical data \cite{carette}. MBPT methods in the finite-field (FF) approach were 
employed by Diercksen and Sadlej \cite{diercksen} to estimate the $\alpha_d$ value of Cl$^-$. Photo-detachment phenomena of Au$^-$ has been rigorously
studied by several experiments \cite{champeau1,champeau2,stapelfeldt}. There are also many practical applications of Au$^-$ in material science and 
chemistry \cite{suga,mcewen}. Both the non-relativistic and approximated relativistic calculations of $\alpha_d$ of Au$^-$ are reported in 
Refs. \cite{kello,schwerdtfeger}. A precise measurement of EA of Au$^-$ has been reported in Ref. \cite{hotop1}, following which a number of
calculations have been carried out to explain the experimental data. Earlier theoretical calculations of energies in both the ions spanned over a 
wide range and disagree with each other \cite{neogrady}. However, recent sophisticated calculations considering higher-level excitations and
higher-order relativistic corrections show excellent agreement with the measurements \cite{nist,pasteka}.

\section{Basic formalism}

The electronic configurations of Cl and Au are $[2p^6]3s^23p^5$ and $[5p^6]4f^{14}5d^{10}6s^1$, respectively. This implies that the electronic 
configurations of Cl$^-$ and Au$^-$ are the $[3p^6]$ and $[5p^6]4f^{14}5d^{10}6s^2$ closed-shell configurations, respectively. The ground state 
$|\Psi_0^{(0)} \rangle$ and its energy $E_0^{(0)}$ due to atomic Hamiltonian ($H_{at}$) without considering any external interaction can be obtained 
by solving the equation
\begin{eqnarray}
 H_{at} |\Psi_0^{(0)} \rangle = E_0^{(0)} |\Psi_0^{(0)} \rangle ,
\end{eqnarray}
where we consider $H_{at}$ as sum of the Dirac Hamiltonian, nuclear potential, and Coulomb repulsion potential ($V_C$) seen by the electrons. Due 
to the two-body nature of $V_C=\frac{1}{2} \sum_{i,j}\frac{1}{r_{ij}}$ (in atomic units (a.u.)), an exact solution of the above equation is not 
feasible. Thus, we express $H_{at}=H_0 + V_{res}$ where $H_0=\sum_i h_i$ contains Dirac Hamiltonian, nuclear potential and an effective one-body 
mean-field potential $U_0=\sum_i u_i$ constructed from $V_C$ and the residual part is defined as $V_{res}=V_C-U_0$. We adopt the Dirac-Fock (DF) 
method to define $U_0$. In this method, the approximated ground state wave function $|\Phi_0 \rangle$ and self-consistent Fock (SCF) energy 
(or DF energy $E_0^{DF}$) are obtained by determining the wave functions for the single-particles as
\begin{eqnarray}
 (h_i+u_i)|\phi_i^{(0)}\rangle &=& \epsilon_i^{(0)}|\phi_i^{(0)}\rangle,
 \label{eq1}
\end{eqnarray}
where $|\phi_i^{(0)}\rangle$ is the $i^{th}$ orbital wave function with energy $\epsilon_i^{(0)}$. The Slater determinant of single-particle wave 
functions form $|\Phi_0 \rangle$ and $E_0^{DF}= \sum_i \epsilon_i^{(0)} +\langle \Phi_0 | V_{res} | \Phi_0 \rangle$. The single-particle mean-field 
potential is defined as
\begin{eqnarray}
 u_i|\phi_i^{(0)}(1) \rangle &=& \sum_a^{N_c} \left [ \langle \phi_a^{(0)}(2)| \frac{1}{r_{12}} |\phi_a^{(0)}(2) \rangle | \phi_i^{(0)}(1)\rangle \right. \nonumber \\
 && \left. - \langle \phi_a^{(0)}(2)| \frac{1}{r_{12}} |\phi_i^{(0)} (2)\rangle | \phi_a^{(0)}(1)\rangle \right ] ,
\label{eq2}
\end{eqnarray}
where $N_C$ represents for the number of electrons in the respective negative ions. It should be noted here that all the orbitals see the 
correlation with the appended electron of the respective negative ion in the above formalism. We have calculated nuclear potential for an electron at
the distance $r$ by assuming finite-size nuclear Fermi charge density distribution, given by \cite{hofstadter}
\begin{eqnarray}
 \rho(r)=\frac{\rho_0}{1+e^{(r-c)/a}},
 \label{eq3}
\end{eqnarray}
where $\rho_0$ is the normalization constant, and the parameter $c$ and $a=4 t ~ ln(3)$ are said to be half-charge-radius and skin thickness of the 
atomic nucleus, respectively. The radial components of the DF single-particle wave functions are expanded using Gaussian type orbitals (GTOs), 
defined for a given orbital angular momentum ($l$) symmetry as \cite{boys}
\begin{eqnarray}
 f_l(r) = \sum_k^{N_l} C_k {\cal N}_k r^l e^{-\alpha_0 \beta^{k-1} r^2}, 
 \label{eq4}
\end{eqnarray}
where $N_l$ denotes number of GTOs, $C_k$ corresponds to expansion coefficient, $\alpha_0$ and $\beta$ are arbitrary parameters that are chosen to 
optimize for the finite-size basis functions, and ${\cal N}_k$ is the normalization factor of the respective GTO and defined in Ref. \cite{mohanty}. 

The exact ground state wave function after including the electron correlation effects from the residual interaction $V_{res}$ can be obtained from 
the above mean-field wave function by operating the wave operator $\Omega_0^{(0)}$ as \cite{lindgren}
\begin{eqnarray}
 |\Psi_0^{(0)} \rangle = \Omega_0^{(0)} |\Phi_0 \rangle .
  \label{eq4}
\end{eqnarray}

In the presence of an external electric field $\vec{\mathcal E}$, the wave function of the negative ion due to the total Hamiltonian 
$H=H_{at}+\vec D \cdot \vec{\mathcal E}$ with dipole operator $D=\sum_i d_i$ can be expressed as
\begin{eqnarray}
 |\Psi_0 \rangle = \Omega_0 |\Phi_0 \rangle ,
  \label{eq5}
\end{eqnarray}
where $\Omega_0$ is the wave operator that is responsible for accounting for electron correlation effects and effects due to the electric field. 
For the weak electric field, we can expand the wave function perturbatively as
\begin{eqnarray}
 |\Psi_0 \rangle &=& |\Psi_0^{(0)} \rangle + |\vec{\mathcal E}| |\Psi_0^{(1)} \rangle + \cdots \nonumber \\
                 &=& \left [ \Omega_0^{(0)} + |\vec{\mathcal E}|\Omega_0^{(1)}  + \cdots \right ] |\Phi_0 \rangle 
  \label{eq6}
\end{eqnarray}
so that
\begin{eqnarray}
\Omega_0 = \Omega_0^{(0)} + |\vec{\mathcal E}|\Omega_0^{(1)}  + \cdots .
 \label{eq7}
\end{eqnarray}
Similarly, the modified energy can be expanded as
\begin{eqnarray}
E_0 &=& E_0^{(0)} + |\vec{\mathcal E}| E_0^{(1)} + \frac{1}{2} |\vec{\mathcal E}|^2 E_0^{(2)}\cdots .
  \label{eq8}
\end{eqnarray}
In the above expressions, superscripts 0, 1, etc. denote order of $\vec{\mathcal E}$ in the expansion. The first-order energy shift ($E_0^{(1)}$) in atomic systems
due to the presence of electric-field vanishes owing to spherical symmetry distribution of charges, but the second-order energy shift ($E_0^{(2)}$)
can be given by
\begin{eqnarray}
 E^{(2)}= \frac{1}{2} ~ \alpha_d ~ |\vec{\mathcal E}|^2 .
 \label{eq8}
\end{eqnarray}
This shift can be estimated with the knowledge of $\alpha_d$ for a given value of $\vec{\mathcal E}$. In molecular systems, $\alpha_d$ is estimated 
conveniently using the FF approach. To adopt the FF approach for determining $\alpha_d$ of atomic systems, it requires to exploit the spherical 
symmetrical property. Thus, the previous calculations of $\alpha_d$ of Cl$^-$ and Au$^-$ are estimated in the FF approach by breaking atomic spherical 
symmetry. To determine $\alpha_d$ values of these ions by preserving spherical symmetry, we adopt the perturbative approach by expressing 
as \cite{yashpal3,bijaya2}
\begin{eqnarray}
\alpha_d &=&  2 \frac{\langle \Psi_0^{(0)}|D|\Psi_0^{(1)} \rangle}{ \langle \Psi_0^{(0)}| \Psi_0^{(0)} \rangle } \nonumber \\
         &=&  2 \frac{\langle \Phi_0| \Omega_0^{(0)\dagger} D \Omega_0^{(1)} |\Phi_0 \rangle}{ \langle \Phi_0| \Omega_0^{(0)\dagger} 
         \Omega_0^{(0)}| \Phi_0 \rangle } .
\label{eq9}
\end{eqnarray}
In the following section, we shall be discussing about how to define both the unperturbed and perturbed wave operators in the DF, relativistic MBPT
(RMBPT), relativistic RPA (RRPA) and RCC methods to fathom about the propagation of electron correlation effects from lower- to all-order perturbative
methods in the evaluation of $\alpha_d$ of the undertaken negative ions. 

Now, we proceed to discuss the general procedure to obtain IP by removing the extra electron from the outer most orbital of the negative ions. For 
this purpose, we define the new working reference state as $|\Phi_a \rangle=a_a |\Phi_0 \rangle$, where $a_a$ denotes annihilation of an electron 
from the outermost orbital $|\phi_a\rangle$ of $|\Phi_0 \rangle$. Accordingly, the wave operator due to $H_{at}$ is defined to obtain the exact 
state as \cite{nandy1}
\begin{eqnarray}
 |\Psi_a\rangle = \Omega_a |\Phi_a \rangle .
\end{eqnarray}
By calculating energy difference between this state and $|\Psi_0^{(0)}\rangle$, one can get IP. Below, we discuss the RCC theory in the Fock-space 
formalism to define $\Omega_a$.

\section{Many-body methods}

\subsection{$\alpha_d$ evaluation}

In the $k^{th}$ order (R)MBPT method ((R)MBPT($k$)), the wave operator can be expanded as \cite{yashpal1,lindgren}
\begin{eqnarray}
 \Omega &=& \sum_{m=1}^k \Omega^{(m)} = \sum_{m=1}^k \sum_{i=0}^m \sum_{j=1}^{k-m} \Omega^{(i,j)} ,
\end{eqnarray}
where $i$- orders of $V_{res}$ and $j$- orders of $D$ are incorporated in the expansion. Thus, the wave operators with zeroth- and first-order $D$
in the RMBPT($n$) method are given by
\begin{eqnarray}
 \Omega^{(0)} &=& \sum_{m=0}^k \Omega^{(m,0)} \ \ \ \text{and} \ \ \ 
 \Omega^{(1)} = \sum_{m=0}^{k-1} \Omega^{(m,1)}
\end{eqnarray}
with $\Omega^{(0,0)}=1$, $\Omega^{(1,0)}=0$ and $\Omega^{(0,1)}= \sum_{p,a} \frac{ \langle \phi_p | d | \phi_a \rangle} {\epsilon_p^{(0)} - \epsilon_a^{(0)}}$ 
for all the occupied orbitals denoted by the index $a$ and unoccupied orbitals denoted by the index $p$. This follows the expression to evaluate the 
lowest-order polarizabilities result in the DF method as
\begin{eqnarray}
 \alpha_d  &=& 2 \langle \Phi_0| {\Omega^{(0,0)}}^{\dagger} D \Omega^{(0,1)} |\Phi_0 \rangle \nonumber \\
         &=& 2 \langle \Phi_0| D \Omega^{(0,1)} |\Phi_0 \rangle .
\end{eqnarray}

The amplitudes of the finite-order unperturbed and perturbed wave operators are obtained using the Bloch's equation \cite{lindgren,Kvasnicka}
\begin{eqnarray}
 [\Omega^{(\beta,0)},H_0 ] P &=& Q V_{res} \Omega^{(\beta-1,0)}P  \nonumber \\ && -
 \sum_{m=1 }^{\beta-1} \Omega^{(\beta-m,0)} P V_{res} \Omega^{(m-1,l)}P \ \  \  \  \
\end{eqnarray}
and using the modified Bloch's equation \cite{yashpal1}
\begin{eqnarray}
 [\Omega^{(\beta,1)},H_0 ]P &=& QV_{res} \Omega^{(\beta-1,1)}P + Q D \Omega^{(\beta,0)}P 
\nonumber \\ && - \sum_{m=1 }^{\beta-1} \big ( \Omega^{(\beta-m,1)}
 P V_{res} \Omega^{(m-1,0)}P \nonumber \\ && - \Omega^{(\beta-m,1)}PD \Omega^{(m,0)}P \big ),
\end{eqnarray}
respectively, with the definitions of model space $P= |\Phi_0 \rangle \langle \Phi_0 |$ and orthogonal space $Q=1-P$. 

This follows the expression for $\alpha_d$ in the RMBPT(3) method as \cite{yashpal1}
\begin{eqnarray}
\alpha_d &=& 2 \frac{\sum_{\beta=0}^{2} \langle \Phi_0| {\Omega^{(2-\beta,0)}}^{\dagger} D \Omega^{(\beta,1)} |\Phi_0 \rangle}
{ \sum_{\beta=0}^{2} \langle \Phi_0| {\Omega^{(2-\beta,0)}}^{\dagger} \Omega^{(\beta,0)} |\Phi_0 \rangle} \nonumber \\
       &=& \frac{2}{\cal N} \langle \Phi_0 | [\Omega^{(0,0)}+\Omega^{(1,0)}+\Omega^{(2,0)}]^{\dagger} D \nonumber \\ && \times[\Omega^{(0,1)}+\Omega^{(1,1)}+\Omega^{(2,1)}]|\Phi_0 \rangle \nonumber \\
&=& \frac{2}{\cal N} \langle \Phi_0| D\Omega^{(0,1)} + D\Omega^{(1,1)}+D\Omega^{(2,1)} + {\Omega^{(1,0)}}^{\dagger} D\Omega^{(0,1)}  \nonumber \\ && +
{\Omega^{(1,0)}}^{\dagger} D\Omega^{(1,1)} +{\Omega^{(2,0)}}^{\dagger} D\Omega^{(0,1)}|\Phi_0 \rangle  ,
\label{eq21}
\end{eqnarray} 
with the normalization constant ${\cal N}=\langle \Phi_0| 1 + {\Omega^{(1,0)}}^{\dagger} \Omega^{(0,1)} |\Phi_0 \rangle$. It can be easily followed
that the lowest-order term corresponds to the DF expression and terms containing up to one-order in $V_{res}$ and one $D$ operator will give rise
expression for the RMBPT(2) method.

Now, we move on to RRPA expression by expanding single-particle DF wave function and energy in the presence of external electric field as \cite{yashpal3}
\begin{eqnarray}
  && |\phi_i \rangle = |\phi_i^{(0)} \rangle+ |\vec{\mathcal E}| |\phi_i^{(1)} \rangle + \cdots  \\
 \text{and} \ \ \ &&  \epsilon_i = \epsilon_i^{(0)} + |\vec{\mathcal E}| \epsilon_i^{(1)} + \cdots .
 \label{eq22}
\end{eqnarray}
 Since the single-particle dipole operator $d$ is odd under parity, $\epsilon_i^{(1)}=0$. To obtain the first-order correction to the single- particle 
wave function, the general single-particle equation is expanded by keeping up to linear in $|\vec{\mathcal E}|$ as 
\begin{eqnarray}
 \left (h_i+ |\vec{\mathcal E}| d_i \right ) \left (|\phi_i^{(0)} (1) \rangle 
 + |\vec{\mathcal E}| |\phi_i^{(1)} (1) \rangle \right ) 
  +  \sum_b^{N_c} \left  (\langle \phi_b^{(0)} (2) \right. \nonumber \\  \left. + |\vec{\mathcal E}| \phi_b^{(1)} (2) |  \frac{1}{r_{12}} 
  |\phi_b^{(0)} (2) + |\vec{\mathcal E}| \phi_b^{(1)}(2) \rangle |\phi_i^{(0)} (1) \right. \nonumber \\ 
  \left.   + |\vec{\mathcal E}| \phi_i^{(1)} (1) \rangle   - \langle \phi_b^{(0)} (2) +  |\vec{\mathcal E}| \phi_b^{(1)} (2) |\frac{1}{r_{12}} |\phi_i^{(0)} (2) \right. \nonumber \\ \left.  + 
 |\vec{\mathcal E}| \phi_i^{(1)} (2) \rangle |\phi_b^{(0)} (1) 
  + |\vec{\mathcal E}| \phi_b^{(1)} (1) \rangle \right ) \nonumber \\
  \simeq \epsilon_i^{(0)} \left (|\phi_i^{(0)} (1) \rangle+ |\vec{\mathcal E}| |\phi_i^{(1)} (1) \rangle \right ). \ \ \ \ \ \ \  
\end{eqnarray}
Retaining only linear in $|\vec{\mathcal E}|$ terms from the above expression, it yields
\begin{eqnarray}
 \left (h_i+u_i-\epsilon_i^{(0)} \right ) |\phi_i^{(1)} \rangle= (-d_i -u_i^{(1)})|\phi_i^{(0)}\rangle,  
\end{eqnarray}
where the modified DF potential $u_i^{(1)}$ is given by
\begin{eqnarray}
 u_i^{(1)} |\phi_i^{(0)}(1) \rangle =\sum_b^{N_c} \left (\langle \phi_b^{(0)}(2)|\frac{1}{r_{12}} |\phi_b^{(1)}(2) \rangle |\phi_i^{(0)} (1)\rangle \right. \nonumber \\
\left. -\langle \phi_b^{(0)}(2)|\frac{1}{r_{12}} |\phi_i^{(0)}(2)\rangle |\phi_b^{(1)}(1) \rangle  
+\langle \phi_b^{(1)}(2)|\frac{1}{r_{12}} |\phi_b^{(0)}(2) \rangle \right. \nonumber \\ \left. \times |\phi_i^{(0)}(1) \rangle 
-\langle \phi_b^{(1)}(2)|\frac{1}{r_{12}} |\phi_i^{(0)}(2)\rangle |\phi_b^{(0)} (1)\rangle \right ) . \ \ \ \ \ 
\end{eqnarray}

  Using the completeness principle, we can write
\begin{eqnarray}
 |\phi_i^{(1)} \rangle= \sum_{j \ne i} C_i^j |\phi_j^0\rangle,
\end{eqnarray}
where $C_i^j$s are the expansion coefficients. Thus, it can be expressed as 
\begin{eqnarray}
 \sum_{j \ne i} C_i^j \left (h_j + u_j - \epsilon_j^{(0)} \right ) |\phi_j^{(0)} \rangle= - \left ( d_i + u_i^{(1)} \right ) |\phi_i^{(0)} \rangle. \ \ \ \ \ 
\end{eqnarray}
This is solved self-consistently to obtain the $C_i^j$ coefficients, hence, $|\phi_i^{(1)} \rangle$ to infinity order in Coulomb interaction 
and one order in the dipole operator by considering contributions only from the singly excited determinants from $|\Phi_0 \rangle$. In RRPA, the 
unperturbed wave operator is taken to be $\Omega^{(0,0)}=1$ and the first-order perturbed wave operator is defined using the above expression by
\begin{widetext}
\begin{eqnarray}
\Omega^{(1)} &=&  \Omega_{\text{RPA}} = \sum_{k=0}^{\infty} \sum_{p,a} \Omega_{a \rightarrow p}^{(k, 1)} \nonumber \\
    &=& \Omega_{a \rightarrow p}^{(0, 1)} + \sum_{\beta=1}^{\infty} \sum_{pq,ab}  \left \{ \frac{ \left [\langle \phi_p^{(0)}(1)\phi_b^{(0)}(2) | \frac{1}{r_{12}} | \phi_a^{(0)}(1) \phi_q^{(0)}(2) \rangle 
- \langle \phi_p^{(0)}(1)\phi_b^{(0)}(2) | \frac{1}{r_{12}} | \phi_q^{(0)}(1)\phi_a^{(0)}(2) \rangle \right ] \Omega_{b \rightarrow q}^{(\beta-1,1)} } {\epsilon_p^{(0)} - \epsilon_a^{(0)}}  \right. \nonumber \\ 
&& \left. \ \ \ \ \ \ \ \ \ \ \ \ \ \ \ + \frac{ \Omega_{b \rightarrow q}^{{(\beta-1,1)}^{\dagger}} \left [\langle \phi_p^{(0)}(1)\phi_q^{(0)}(2) | \frac{1}{r_{12}} | \phi_a^{(0)}(1)\phi_b^{(0)}(2) \rangle - 
\langle \phi_p^{(0)}(1)\phi_q^{(0)}(2) | \frac{1}{r_{12}} | \phi_b^{(0)}(1)\phi_a^{(0)}(2) \rangle \right]} 
 {\epsilon_p^{(0)}-\epsilon_a^{(0)}} \right \},
\end{eqnarray} 
\end{widetext}
where $a \rightarrow p$ means replacement of an occupied orbital $|\phi_a\rangle$ from $|\Phi_0 \rangle$ by a virtual orbital $|\phi_p\rangle$ which alternatively refers to a 
singly excited state with respect to $|\Phi_0 \rangle$. It can be understood from the above formulation that the RRPA method picks-up a certain class of 
single excitation configurations by capturing the core-polarization correlation effects to all-orders. Again, contributions included in this 
perturbative approach is equivalent to the orbital relaxation effects that arise at the DF method in the FF approach. 

Using the above wave operator, we evaluate $\alpha_d$ in RRPA as
\begin{eqnarray}
 \alpha_d  &=& 2 \langle \Phi_0| {\Omega^{(0,0)}}^{\dagger} D \Omega^{(1)} |\Phi_0 \rangle \nonumber \\
         &=& 2 \langle \Phi_0| D \Omega_{\text{RPA}} |\Phi_0 \rangle .
\end{eqnarray}

In the RCC method, the wave operator including the external perturbation has the form 
\begin{eqnarray}
\Omega &=& e^T , 
\end{eqnarray} 
where $T$ is known as the excitation operator that is responsible to take care of electron correlation effects from the reference state $|\Phi_0 \rangle$ 
due to $V_{res}$ and $D$ operators. By expanding $T$ in $|\vec{\mathcal E}|$, and keeping zeroth and linear terms gives us \cite{yashpal1,yashpal2,yashpal3,bijaya3}
\begin{eqnarray}
\Omega^{(0)} &=& e^{T^{(0)}}  \ \ \ \ \text{and} \ \ \ \ \Omega^{(1)}= e^{T^{(0)}} T^{(1)} ,
\end{eqnarray} 
respectively. The amplitudes of the excitation operator $T^{(0)}$ and energy $E^{(0)}$ are determined by projecting the excited determinants
as \cite{yashpal1,bijaya3}
\begin{eqnarray}
 \langle \Phi_{\tau}|\overline{H_{at}}|\Phi_0\rangle&=& E_0^{(0)} \delta_{\tau,0},
 \end{eqnarray}
where notation $\overline{O}=(Oe^{T^{(0)}})_c$ is used with subscript $c$ means connected terms and $|\Phi_{\tau}\rangle$ means excited Slater 
determinants with respect to $|\Phi_0\rangle$. Similarly, the amplitudes of the excitation $T^{(0)}$ operator (note that energy $E^{(1)}=0$)
are obtained by solving the equation
\begin{eqnarray}
\langle \Phi_{\tau}|\overline{H_{at}}T^{(1)} + \overline{D}|\Phi_0\rangle =0 .
\end{eqnarray}
In our calculations, we consider only the singles and doubles excited configurations in the RCC theory (RCCSD method) by denoting $\tau \equiv 1$ and 2, 
respectively, and the RCC operators as 
\begin{eqnarray}
T^{(0)} &=& T_1^{(0)} + T_2^{(0)} \ \ \ \ \text{and} \ \ \ \ 
T^{(1)} = T_1^{(1)} + T_2^{(1)} .
\end{eqnarray}

In the RCC theory, the $\alpha_d$ determining expression is given by \cite{yashpal4,bijaya4}
\begin{eqnarray}
 \alpha_d &=&2 \frac{\langle\Phi_0 | \Omega^{(0) \dagger} D \Omega^{(1)} | \Phi_0 \rangle }
                  {\langle\Phi_0 | \Omega^{(0) \dagger} \Omega^{(0)} | \Phi_0 \rangle } \nonumber \\
        &=&2 \frac{\langle\Phi_0 | e^{T^{(0) \dagger}} D e^{T^{(0)}} T^{(1)} | \Phi_0 \rangle }
                  {\langle\Phi_0 | e^{T^{(0) \dagger}} e^{T^{(0)}} | \Phi_0 \rangle } \nonumber \\
        &=&2 \langle\Phi_0 |(\overbrace{D^{(0)}} T^{(1)})_c|\Phi_0 \rangle,
  \label{prcc}
\end{eqnarray}
where $\overbrace{D^{(0)}} = e^{T^{\dagger{(0)}}}De^{T^{(0)}}$ is a non-truncating series. The above expression is derived from the property 
evaluation expression given by Refs. \cite{pal1,pal2}. We have adopted an iterative procedure to take into 
accounting contributions from this non-terminating series self-consistently as described in our earlier works on $\alpha_d$ calculations 
in the closed-shell atoms \cite{bijaya2,bijaya3}.

\subsection{IP evaluation}

In the Fock-space approach, the wave operator describing removal of an electron from orbital $|\phi_a\rangle$ of $|\Phi_0 \rangle$ is defined 
in the RCC theory by \cite{debashis,lindgn,nandy1,nandy2,nandy3}
\begin{eqnarray}
\Omega_a = e^{T^{(0)}} (1+ R_a),
\end{eqnarray}
where $R_a$ is another RCC operator introduced to take care of the extra correlation effects that was included through the detached electron. Then, 
the energy ($E_a$) of the product state and amplitudes of the $R_a$ operator is obtained by solving
\begin{eqnarray}
\langle \Phi_{\eta}| \overline{H_{at}} R_a +\overline{H_{at}} |\Phi_a\rangle &=& \langle \Phi_{\eta}| \left [ \delta_{\eta,a} +
R_a \right ] |\Phi_a\rangle E_a ,
\label{ccen}
\end{eqnarray}
where $|\Phi_{\eta} \rangle$ is designated as the excited configuration determinants from $|\Phi_a\rangle$ for the $R_a$ amplitude determination else 
it corresponds to $|\Phi_a\rangle$ to estimate $E_a$. Hence, the IP of the electron removed from $|\phi_a\rangle$ is obtained by taking the difference
as $\Delta E_a = E_0^{(0)} - E_a$. Here, we have also considered the RCCSD method approximation by considering singles and doubles excited 
configurations for $|\Phi_{\eta} \rangle$.

\subsection{Atomic Hamiltonian}

The starting point of our calculation is the Dirac-Coulomb (DC) Hamiltonian \cite{Dirac} representing the leading order contributions to $H_{em}$ to calculate
the zeroth-order wave functions and energies which in atomic units (a.u.) is given by
\begin{eqnarray}\label{eq:DHB}
H^{DC} &=& \sum_i \left [c\mbox{\boldmath$\alpha$}_i\cdot \textbf{p}_i+(\beta_i-1)c^2+V_n(r_i)\right] +\sum_{i,j>i}\frac{1}{r_{ij}}, \ \ \ \
\end{eqnarray}
where $\mbox{\boldmath$\alpha$}$ and $\beta$ are the usual Dirac matrices, $\textbf{p}$ is the single particle momentum operator, $V_n(r)$ 
denotes the nuclear potential, and $\sum_{i,j}\frac{1}{r_{ij}}$ represents the Coulomb potential between the electrons located at the 
$i^{th}$ and $j^{th}$ positions. It should be noted that the above Hamiltonian is scaled with respect to the rest mass energies of electrons.
Contributions from the Breit interaction \cite{breit} to $H_{em}$ is determined by including the following potential 
\begin{eqnarray}\label{eq:DHB}
V^B &=& - \sum_{j>i}\frac{[\mbox{\boldmath$\alpha$}_i\cdot\mbox{\boldmath$\alpha$}_j+
(\mbox{\boldmath$\alpha$}_i\cdot\mathbf{\hat{r}_{ij}})(\mbox{\boldmath$\alpha$}_j\cdot\mathbf{\hat{r}_{ij}})]}{2r_{ij}} ,
\end{eqnarray}
where $\mathbf{\hat{r}_{ij}}$ is the unit vector along $\mathbf{r_{ij}}$.

Contributions from the QED effects to $H_{em}$ are estimated by considering the lower-order vacuum polarization (VP) interaction ($V_{VP}$) 
and the self-energy (SE) interactions ($V_{SE}$). We account for $V_{VP}$ through the Uehling \cite{Uehl} 
and Wichmann-Kroll \cite{Wichmann} potentials ($V_{VP}=V^{Uehl} + V^{WK}$), given by
\begin{eqnarray}
 \label{eq:uehl}
V^{Uehl}&=&- \frac{2}{3} \sum_i \frac{\alpha_e^2 }{r_i} \int_0^{\infty} dx \ x \ \rho(x)\int_1^{\infty}dt \sqrt{t^2-1} \nonumber \\
&& \times\left(\frac{1}{t^3}+\frac{1}{2t^5}\right)  \left [ e^{-2ct|r_i-x|} - e^{-2ct(r_i+x)} \right ]\ \ 
\end{eqnarray}
and
\begin{eqnarray}
 V^{WK} = \sum_i \frac{0.368 Z^2}{9 \pi c^3 (1+(1.62 c r_i )^4) } \rho(r_i),
\end{eqnarray}
respectively, where $\alpha_e$ is the fine structure constant.

The SE contribution $V_{SE}$ is estimated by including two parts \cite{Flambaum}
\begin{eqnarray}
V_{SE}^{ef}&=&  A_l \sum_i \frac{2 \pi Z \alpha_e^3 }{r_i} I_1^{ef}(r_i) - B_l \sum_i \frac{\alpha_e }{ r_i} I_2^{ef}(r_i) \ \ \
\end{eqnarray}
known as the effective electric form factor part and
\begin{eqnarray}
V_{SE}^{mg}&=& - \sum_k \frac{i\alpha_e^3}{4} \mbox{\boldmath$\gamma$} \cdot \mbox{\boldmath$\nabla$}_k \frac{1}{r_k} \int_0^{\infty} dx \ x \ \rho(x)
\int_1^{\infty} dt \frac{1}{t^3 \sqrt{t^2-1}} \nonumber \\
\times && \left [ e^{-2ct|r_k-x|} - e^{-2ct(r_k+x)} - 2ct \left (r_k+x-|r_k-x| \right ) \right ], \nonumber \\
\end{eqnarray}
known as the effective magnetic form factor part. In the above expressions, we use \cite{Ginges} 
\begin{eqnarray}
A_l= \begin{cases} 0.074+0.35Z \alpha_e \ \text{for} \ l=0,1 \\  0.056+0.05 Z \alpha_e + 0.195 Z^2 \alpha_e^2 \ \text{for} \ l=2  , \end{cases}
\end{eqnarray}
and
\begin{eqnarray}
B_l = \begin{cases} 1.071-1.97y^2 -2.128 y^3+0.169 y^4  \ \text{for} \ l=0,1 \\
     0 \ \text{for} \ l \ge 2 .  \end{cases}   
\end{eqnarray}
The integrals are given by
\begin{eqnarray}
I_1^{ef}(r) =  \int_0^{\infty} dx \ x \ \rho(x) [ (Z |r-x|+1) e^{-Z|r-x|} \nonumber \\  - (Z(r+x)+1) e^{-2ct(r+x)}  ] \ \ \ \ \ \ 
\end{eqnarray}
and
\begin{eqnarray}
 I_2^{ef}(r) &=& \int_0^{\infty} dx \ x \ \rho(x)  \int^{\infty}_1 dt \frac{1}{\sqrt{t^2-1}} \bigg \{ \left( 1-\frac{1}{2t^2} \right ) \nonumber \\
&\times& \left [ \ln(t^2-1)+4 \ln \left ( \frac{1}{Z \alpha_e} +\frac{1}{2} \right ) \right ]-\frac{3}{2}+\frac{1}{t^2} \big \}\nonumber \\
&\times& \{ \frac{\alpha_e}{t} \left [ e^{-2ct|r-x|} - e^{-2ct(r+x)} \right ] +2 r_A e^{2 r_A ct } \nonumber \\
&\times& \left [ E_1 (2ct (|r-x|+r_A)) - E_1 (2ct (r+x+r_A)) \right ] \bigg \} \nonumber \\
\end{eqnarray}
with the orbital quantum number $l$ of the system, $y=(Z-80)\alpha_e$, $r_A= 0.07 Z^2 \alpha_e^3$, and the exponential integral $E_1(r) = 
\int_r^{\infty} ds e^{-s}/s$.

\begin{table}[t]
\caption{Calculated $\alpha_d$ values (in a.u.) of Cl$^-$ by Diercksen and Sadlej \cite{diercksen} in the FF approach. Results from basis I 
(14$s$11$p$5$d$) and basis II (17$s$13$p$5$d$) are reported at different levels of approximations in MBPT method. Results from basis I after 
including all core orbitals (All), and freezing core orbitals from the K and L shells (Frozen) are also given.}
\begin{ruledtabular}
\begin{tabular}{lcccc} 
   Method  & \multicolumn{3}{c}{Basis I}  & Basis II  \\
 \cline{2-4}  \\ 
           &  All & Frozen K &  Frozen K$+$L &   All \\  
\hline \\
   HF                         &  31.45  &  31.45  & 31.45  & 31.56 \\
   MBPT$^{\rm D}$(2)          &  37.06  &  37.07  & 37.00  & 37.25 \\
   MBPT$^{\rm D}$(3)          &  29.91  &  29.90  & 29.93  & 29.99  \\
   MBPT$^{\rm SD}$(4)         &  36.71  &  36.71  & 36.62  & 36.87 \\
  MBPT$^{\rm SD[1/1]}$(4)     &  38.82  &  39.04  & 37.52  & 39.20 \\
  MBPT$^{\rm SD}$($\tilde{4}$)&  35.47  &  35.44  & 35.22  & 35.63   \\
\end{tabular}
\end{ruledtabular}
\label{tab10}
\end{table}

\section{Results and Discussion}

 We would like to first discuss briefly about the previous calculated values of $\alpha_d$ in both the Cl$^-$ and Au$^-$ ions to understand the need 
of doing new theoretical results by including correlations effects among all the electrons rigorously through relativistic many-body methods. For this 
purpose, we give the results for Cl$^-$ in Table \ref{tab10} from the only one calculation reported by Diercksen and Sadlej \cite{diercksen} by 
incorporating electron correlation effects using the non-relativistic (NR) MBPT($k$) methods with $k=2,3$ and 4 denoting the order of residual 
Coulomb interaction. To demonstrate roles of the core-orbitals, they had analyzed results by considering all core electrons and freezing core-orbitals
from the K- and L-shells by using a set of basis as 14$s$11$p$5$d$ (basis I). Then, they had used a slightly larger basis (basis II) by appending a 
few more high-lying $s$, $p$ and $d$ orbitals to basis I as 17$s$13$p$5$d$ to show contributions from the high-lying orbitals. The differences between 
results from both the basis functions were found to be insignificant. There are three major limitations of this calculation: (i) It uses NR theory,
however, later theoretical studies have exhibited quite large relativistic effects in the determination of $\alpha_d$ values of the negative ions 
\cite{kello,schwerdtfeger,alkal}. 
(ii) It considers only either double excitations (denoted by MBPT$^{\rm D}$) or single and double excitations (denoted by MBPT$^{\rm SD}$) excitations
even in the MBPT methods. (iii) It has completely ignored correlation contributions from the higher-symmetry orbitals such as $f$, $g$, etc.. Since 
$\alpha_d$ involves E1 operator, whose matrix element is directly proportional to radial distance, contributions from the higher angular momentum 
orbitals cannot be completely ignored. Nonetheless, a recommended value of $\alpha_d$ of Cl$^-$ was reported as 37.5 a.u. by Diercksen and Sadlej 
after analyzing correlation energy trends and taking into account corrections from the Pade approximants though the MBPT$^{\rm SD}[1/1]$(4) 
approximation and considering invariant fourth-order (denoted by $\tilde{4}$) contribution through MBPT$^{\rm SD}$($\tilde{4}$) approximation 
\cite{diercksen}. This recommended value was, however, not close to any of the their results obtained using the first-principle calculations. 
Therefore, it is necessary to perform more accurate calculation of $\alpha_d$ of Cl$^-$ to ascertain its value by incorporating correlation effects 
among all electrons more rigorously in the RCC theory.  

\begin{table}[t]
\caption{Calculated $\alpha_d$ values (in a.u.) of Au$^-$ by Schwerdtfeger and Bowmaker \cite{schwerdtfeger} using the MBPT, QCISD and QCISD(T) methods.
Results obtained using the non-relativistic and $j$-averaged relativistic pseudopotentials are given separately. Number of active orbitals considered 
in the calculations are denoted by $N$.}
\begin{ruledtabular}
\begin{tabular}{lcc} 
   Method  &       $N$      &      Result \\
\hline \\
\multicolumn{3}{c}{\underline{Non-relativistic}} \\
   HF          &       &   660.07   \\
   MBPT(2)     &   20  &    27.47   \\
   MBPT(3)     &   20  &   199.02   \\
   MBPT(4)     &   20  &   279.59   \\
   QCISD       &   20  &   399.18   \\
   QCISD(T)    &   20  &   257.12   \\
   QCISD(T)/$f$  &   20  &   362.87   \\
   & & \\
\multicolumn{3}{c}{\underline{$j$-averaged relativistic}} \\
    DF         &       &  204.95   \\
   MBPT(2)     &  20   &   3.04   \\
   MBPT(3)     &  20   &   62.29   \\
   MBPT(4)     &  20   &   60.60   \\
   QCISD       &  20   &  118.71   \\
   QCISD(T)    &  20   &   96.03   \\
   QCISD(T)/$f$  &  20   &  121.88   \\
\end{tabular}
\end{ruledtabular}
 Here, ``$f$" denotes for ``without metal $f$ functions".
\label{tab11}
\end{table}

Now we turn to discuss about the earlier calculations of $\alpha_d$ for Au$^-$. Compared to the Cl$^-$ ion, there are two rigorous calculations available for 
$\alpha_d$ of Au$^-$. Schwerdtfeger and Bowmaker \cite{schwerdtfeger} had carried out calculation of this quantity by using pseudoptentials 
in the NR theory framework and using spin-orbit ($j$)-averaged relativistic approach. They had applied MBPT($k$) approximations, with $k=2,3$ and 4, 
and quadratic configuration interaction method with singles and doubles approximation (QCISD method) and QCISD method with partial triple excitations 
(QCISD(T) method) in the FF approach. Results from these methods, both in the NR and relativistic frameworks, are given in Table \ref{tab11}. It can 
be seen that there are huge differences among the results from the NR and relativistic calculations at the same level of approximation in the 
many-body methods. These calculations have also several limitations such as they use pseudopotentials instead of the HF potentials, relativistic 
effects are approximated to $j$-averaged approach which cannot consider the exact relativistic effects, and only $N=20$ number of active electrons
were allowed to correlate out of total 80 electrons of Au$^-$. Therefore, contributions from the correlation effects from the remaining 60 electrons 
need to be investigated. Moreover, there are very large differences between results at various approximations were seen (without showing any signature of 
convergence of result with the higher-order contributions). The difference in the results with and without considering metal function $f$ in the 
QICSD(T) method was also found to be quite large. Since it does not provide a recommended value, one cannot be very confident to use those results 
in any of the applications. 

\begin{table}[t]
\caption{Calculated $\alpha_d$ values (in a.u.) of Au$^-$ by Kell\"o {\it et al.} \cite{kello} using NR, MVD and DKnp Hamiltonians. Results using 
the uncontracted basis with DKnp Hamiltonian are given as DKnp$^*$. Calculations are carried out in the MBPT(2), CCSD and CCSD(T) approximations  
by considering three different number of active orbitals $N$.}
\begin{ruledtabular}
\begin{tabular}{lcrrrr} 
   Method  &     $N$      &    NR   &  MVD   &  DKnp  & DKnp$^*$    \\
\hline \\
    SCF         &       &   630    &  101   &  193   &  195  \\
   MBPT(2)      &  12   &  62     & $-106$ &   13   &   15  \\
                &  18   & $-4$    & $-136$ & $-12$  & $-10$ \\
                &  20   &   6     &   138  &  -     &  -    \\
    CCSD        &  12   &   318   &  48    &  109   &  112  \\
                &  18   &   303   &  44    &  105   &  108  \\
                &  20   &   307   &  45    &  -     &  - \\
 CCSD(T)        &  12   &   267   &  27    &  97    &  98   \\
                &  18   &   249   &  21    &  92    &  92   \\
                &  20   &   256   &  23    &  -     &  -   \\
\end{tabular}
\end{ruledtabular}
\label{tab12}
\end{table}

Later, Kell\"o {\it et al.} have made a systematic analysis of $\alpha_d$ of the  negative ions of the coinage metal atoms including 
Au$^-$ \cite{kello}. These calculations were also carried out in the FF approach and they had investigated relativistic effects more judiciously
by analyzing results from the the quasi-relativistic corrections from the mass-velocity and Darwin (MVD) corrections over NR results and DKnp 
Hamiltonian. They had used NpPolMe basis functions without and with fully uncontracted orbitals and demonstrated roles of electron correlation 
effects by applying MBPT(2) method, and CC method with singles and doubles approximation (CCSD) and the CCSD method with partial triples approximation
(CCSD(T)) systematically. A qualitative agreement between the calculations by Kell\"o {\it et al.} and that of Schwerdtfeger and Bowmaker was 
observed. Their finding reveals that the quasi-relativistic corrections from the MVD terms bring down the results by more than half to the NR results 
in the HF method as well as in the CCSD and CCSD(T) methods. They also showed that the results from the DKnp Hamiltonian are very different from
their NR$+$MVD results. However, electron correlation effects only from a fewer electrons were included in their calculations. They had considered 
active orbitals as $N=12$, 18 and 20 for the NR calculations, but they used only $N=12$ and 18 for the relativistic calculations using the DKnp 
Hamiltonian. In fact, their basis functions were also quite small, which considered only $3s3p1d1f$ diffuse functions over the NpPolMe basis functions
\cite{kello}. Results obtained by Kell\"o {\it et al.} at different levels of approximations are given in Table \ref{tab12}. This calculation also 
does not provide any recommended $\alpha_d$ value of Au$^-$ and there was no estimate of uncertainty. 

\begin{table}[t]
\caption{Our calculated $\alpha_d$ values (in a.u.) of Cl$^-$ and Au$^-$ from different relativistic methods using the DC Hamiltonian. Estimated 
corrections from higher-order effects and uncertainties are also given. The final recommended values are given after accounting for possible 
uncertainties and results from the even-parity channel are shown with * mark.}
\begin{ruledtabular}
\begin{tabular}{lcrcr} 
 Method      &  \multicolumn{2}{c}{Cl$^-$}  &   \multicolumn{2}{c}{Au$^-$}  \\
 \cline{2-3} \cline{4-5} \\
             &    $N$   & Result &  $N$   & Result \\ 
 \hline \\
 \multicolumn{5}{c}{Results using DC Hamiltonian} \\
 DF          &   18   & 25.66  &  80  & 122.64   \\
 RMBPT(2)    &   18   & 27.79  &  80  & 138.88   \\
 RMBPT(3)    &   18   & 20.45  &  80  &  60.69   \\
 RRPA        &   18   & 31.71  &  80  & 194.61   \\
 RCCSD*      &   18   & 33.64  &  80  &  95.66   \\    
 RCCSD       &   18   & 35.68  &  80  &  94.30   \\
 \multicolumn{5}{c}{Corrections} \\
 Triples     & 18   &   0.42   &  80  & $-2.28$  \\
 Breit       & 18   &  0.02    &  80  & 1.41    \\
 QED         & 18   & $-0.01$  &  80  &  3.32   \\
 \hline \\
  Final &    &  35(1) &  &  97(3) \\
\end{tabular}
\end{ruledtabular}
\label{tab1}
\end{table}

To improve the calculations of $\alpha_d$ in Cl$^-$ and Au$^-$, we have considered the RCC theory in the perturbative approach by using four-component
relativistic Hamiltonian. We have used 40 GTOs for each angular momentum symmetry up to $l=4$ (i.e. $g$-symmetry) for the generation of single particle orbitals. All the 
electrons are correlated up to principal quantum number $n=20$ virtual orbitals to carry out calculations using the RMBPT, RRPA and RCCSD methods. The $\alpha_d$ 
values are discussed and compared with the previous works first, then we present the IP results. In Table \ref{tab1}, we give the $\alpha_d$ values of
both the Cl$^-$ and Au$^-$ negative ions that are obtained by using DC Hamiltonian in the relativistic many-body methods described in the previous 
section and adding corrections from the neglected effects. As can be seen, the trends in the results from the DF to RCCSD methods using the DC
Hamiltonian are quite different in both the ions. As described in Ref. \cite{rajat}, the even-parity channel multipoles usually contribute 
predominantly to the electron correlation effects in the RCC calculations. To demonstrate their roles here, we have also presented results considering
only the even-parity multipoles in the RCCSD method (marked as RCCSD* to distinguish from the all-parity channel calculations). As can be seen 
there are significant differences in the results from both the channels. It is worth mentioning that it is possible to evaluate results from both 
the channels only when the spherical coordinate system is used to describe the atomic wave functions. In Cl$^-$, the DF method gives a lower value and
RMBPT(2) increases it to a larger value. Then, the RMBPT(3) method brings it down and makes its value lower than that of the DF value. After that the
RRPA makes it larger than the RMBPT(2) value and then, the RCCSD method gives the largest value. This trend is almost similar to the calculation of 
$\alpha_d$ in the isoelectronic atom Ar of Cl$^-$, however, the $\alpha_d$ value of Cl$^-$ is found to be about three times larger than the value of 
Ar \cite{yashpal1}. Moreover, variation in the results from lower- to higher-order methods are not abrupt compared to the FF approach discussed above.
Since its previous calculation by Diercksen and Sadlej \cite{diercksen} was performed in the FF approach through the molecular code, their mean-field
result using the Hartree-Fock (HF) method is equivalent to RPA approximation in the perturbative approach. This is why comparison between our RRPA 
value and the HF value of Ref. \cite{diercksen} shows a very good agreement. Further, Diercksen and Sadlej had employed NR method in contrast to our
relativistic calculation. So good agreement between our RRPA result with the above HF value of FF approach indicates that the relativistic effects 
play less important roles in the determination of $\alpha_d$ of Cl$^-$.   

We have also given the calculated $\alpha_d$ values of Au$^-$ from the considered relativistic many-body methods in Table \ref{tab1}. The trends in 
these results from the DF to RMBPT(3) methods look analogous to the calculations in Cl$^-$, but the RRPA result is found to be much higher than the 
RCCSD value in this case. This trend has similarity with the calculation of $\alpha_d$ of the isoelectronic atom Hg of Au$^-$, but the 
result of Au$^-$ is about three times larger than Hg \cite{yashpal4,bijaya4}.  Our RRPA value is found to be in good agreement with the HF results of 
the earlier calculations using the $j$-averaged pseudopotential \cite{schwerdtfeger} and DK Hamiltonian \cite{kello}; better agreement with the later
one. We have also observed large differences in the results from the RMBPT(2) and RMBPT(3) methods like the previous studies. Though there are large
differences between our RCCSD results with the earlier discussed CCSD and QCISD results are observed, we find that our RCCSD value agrees quite
well with the CCSD(T) and QCISD(T) results. In a recent study on Cd atom \cite{bijaya5}, we had observed that dipole polarizability value converges 
faster in the perturbative approach than FF approach with respect to level of higher excitations. This may have been the reason for the good agreement
between our RCCSD result obtained in the perturbative approach and CCSD(T)/QCISD(T) results than the CCSD/QCISD results of the FF approach.

  From the comparison between the $\alpha_d$ results of both the negative ions with their isoelectronic neutral atoms, it appears to us that these 
quantities change drastically when there is imbalance between nuclear and electronic charges. We have quoted the final $\alpha_d$ values from our 
calculations by adding the estimated corrections from the Breit interaction, quantum electrodynamics (QED) effects and triple excitations. We have 
used RRPA to estimate the Breit and QED contributions, whereas the triple excitation contributions are estimated by defining the following excitation 
operators in the perturbative approach \cite{Kaldor,Watts}
{\small{
\begin{eqnarray}
 T_3^{(0),pert}= \frac{1}{(3!)^2}\sum_{abc,pqr}  \frac{ ( H_{at} T_2^{(0)})_{abc}^{pqr} }{\epsilon_a^{(0)} + \epsilon_b^{(0)} +\epsilon_c^{(0)}-\epsilon_p^{(0)} 
 -\epsilon_q^{(0)} -\epsilon_r^{(0)}} \ \ \ \ \
 \label{eqt30}
\end{eqnarray}
}}
and
{\small{
\begin{eqnarray}
 T_3^{(1),pert}= \frac{1}{(3!)^2}\sum_{abc,pqr}  \frac{ ( H_{at} T_2^{(1)})_{abc}^{pqr} }{\epsilon_a^{(0)} + \epsilon_b^{(0)}+\epsilon_c^{(0)}-
 \epsilon_p^{(0)} -\epsilon_q^{(0)} -\epsilon_r^{(0)}} , \ \ \ \ \
 \label{eqt31}
\end{eqnarray}
}}
where $a,b,c$ and $p,q,r$ subscripts denoting for the occupied and unoccupied orbitals, respectively, and subscripts 0(1) correspond to (un)perturbed 
excitation operators. These operators are directly used in Eq. (\ref{prcc}) as part of the $T^{(0)}$ and $T^{(1)}$ RCC operators to estimate the 
approximated contributions due to the triple excitations. We have also estimated uncertainties to the final values due to use of finite size basis 
functions. We obtain the final $\alpha_d$ values of Cl$^-$ and Au$^-$ as 35(1) a.u. and 97(3) a.u. respectively.

\begin{table}[t]
\caption{Contributions to $\alpha_d$ values (in a.u.) of Cl$^-$ and Au$^-$ ions from different RCC terms. Terms that are not shown explicitly, their 
contributions are quoted together as `Others'.}
\begin{ruledtabular}
\begin{tabular}{lrr}
  RCC term                  & Cl$^-$  & Au$^-$  \\
\hline \\
$DT_1^{(1)}$                &  36.51  & 124.96  \\
$T_1^{(0)\dagger}DT_1^{(1)}$&   0.34  & $-19.10$ \\
$T_2^{(0)\dagger}DT_1^{(1)}$& $-2.82$ & $-13.59$ \\
$T_1^{(0)\dagger}DT_2^{(1)}$&   0.04  & 1.38 \\
$T_2^{(0)\dagger}DT_2^{(1)}$&   1.99  & 4.96 \\
Others                      & $-0.38$ & $-4.31$ \\
\end{tabular}
\end{ruledtabular}
\label{tab2}
\end{table}

We would also like to discuss the trends of electron correlation effects by comparing individual RCC term contributions from the DC Hamiltonian to $\alpha_d$ of both the 
ions. In Table \ref{tab2}, we give the contributions from various RCC terms to $\alpha_d$ of the Cl$^-$ and Au$^-$ negative ions. In both the cases, 
$DT_1^{(1)}$ contributes the highest as it contains the DF value and core-polarization effects to all-orders. These contributions are different than 
RRPA results as here the core-polarization effects are also coupled with the pair-correlation correlations, and there are also additional non-RPA
contributions arising through the non-linear RCC terms \cite{bijaya3}. We find that $T_1^{(0)\dagger}DT_1^{(1)}$ contributes negligibly small in 
Cl$^-$, but it contributes significantly to Au$^-$. Contributions from $T_2^{(0)\dagger}DT_1^{(1)}$ are found to be important in both the ions. The 
reason for this finding is that the lowest-order correlation effects from the unperturbed and perturbed RCC operators come through $T_2^{(0)}$ and
$T_1^{(1)}$, respectively. The remaining terms are found to be less important in the determination of $\alpha_d$ values of both the ions.

\begin{table}[t]
\caption{IP values (in eV) of both Cl$^-$ and Au$^-$ negative ions from various calculations by approximating many-body methods at different levels.
Main results using the DC Hamiltonian and corrections due to the higher-order effects are given separately. Uncertainties are quoted along with 
the final results. The experimental results and previously calculated values are also listed. Results from even-parity channel are shown with * 
mark. We have used conversion factor 1 cm$^{-1}$= 0.00012397788 eV to mention all the results in the same units.}
\begin{ruledtabular}
\begin{tabular}{lccc} 
 Method      &    Cl$^-$   & Au$^-$       & Reference \\
 \hline \\
  \multicolumn{4}{c}{\underline{From the DC Hamiltonian}} \\
 DHF         &   4.027     &  1.177      & This work \\
 RMBPT(2)    &   3.070     &  2.297      & This work  \\
 RCCSD*      &   3.786     &  2.286      & This work \\
 RCCSD       &   3.735     &  2.232      & This work \\
 \multicolumn{4}{c}{\underline{Corrections}} \\
 Breit          &   0.002   &   $-0.007$    &  \\
 QED            &   0.001 &    0.015   &  \\
 Triples        & $-0.113$  &  $0.094$   &  \\
 \hline \\
 Final      & 3.63(5) & 2.33(5) & This work \\
 \hline \\
  \multicolumn{4}{c}{\underline{Other works}} \\
 CCSD(T)     &             & 2.229        & Ref. \cite{neogrady} \\ 
 QCISD(T)    &             & 2.073        & Ref. \cite{schwerdtfeger} \\
 RCCSD       &             & 2.269        & Ref. \cite{eliav}  \\ 
 DC-CCSDTQP  &             & 2.3072       & Ref. \cite{pasteka} \\
 $+$Breit$+$QED            &              &   \\
 CCSD(T)     &  3.509     &              &  Ref. \cite{nist} \\
 Experiment  & 3.6125(6)   &    & Ref. \cite{trainham} \\ 
 Experiment  & 3.6125(3)   &    & Ref. \cite{berzinsh} \\
 Experiment  &             & 2.30863(3)  & Ref. \cite{hotop2} \\ 
\end{tabular}
\end{ruledtabular}
\label{tab3}
\end{table}

Now, we turn to discuss the IP values of Cl$^-$ and Au$^-$ ions. We give these values in Table \ref{tab3} from the DF, RMBPT(2), RCCSD* and RCCSD methods 
using the DC Hamiltonian. It can be noted that the extra electron present in the $3p_{3/2}$ outer orbital in Cl$^-$, whereas it is in the $6s_{1/2}$ 
orbital in Au$^-$. Thus, the outer electron in Cl$^-$ is more tightly bound than Au$^-$. As can be seen in the above table, the correlation trends are
different in both the cases because of the above said reason. The DF result in Cl$^-$ is higher than the RCCSD result, where the DF value is slightly 
higher than half of the RCCSD value in Au$^-$. The RMBPT(2) method gives relatively smaller correlation contributions to the determination of IP in 
Cl$^-$, whereas it gives comparatively larger correlation contributions in Au$^-$. We have also estimated corrections from the Breit and QED interactions 
using the RMBPT(2) method and quoted them in the above table. In this case also we find that there are large differences between the results from the 
RCCSD* and RCCSD methods, and the results from the all-parity channel are more reliable. To estimate the corrections from the triple excitations, 
we construct a perturbative valence triple excitation as
{\small{
\begin{eqnarray}
 R_{3a}^{pert}= \frac{1}{(2!)^2}\sum_{abc,pqr}  \frac{ ( \overline{H_{at}} R_a)_{abc}^{pqr} }{\epsilon_a^{(0)} + \epsilon_b^{(0)}+\epsilon_c^{(0)}-
 \epsilon_p^{(0)} -\epsilon_q^{(0)} -\epsilon_r^{(0)}} . \ \ \ \ \
 \label{eqt32}
\end{eqnarray}
}}
This is used only in the energy evaluating expression of Eq. (\ref{ccen}) after obtaining amplitudes of the RCCSD operators and the estimated 
contributions are given in Table \ref{tab3}. We have also estimated uncertainties from the finite size basis functions to the RCCSD values using 
the DC Hamiltonian. After taking into account all these contributions, we obtain the final IP values of Cl$^-$ and Au$^-$ as 3.63(5) eV and 2.33(5) 
eV respectively. We also compare our results with the available calculations and experimental values. Two precise experimental values of IP for 
Cl$^-$ have been reported in Refs. \cite{trainham,berzinsh} and our result agrees within the error bars of the experimental values. A list of data 
for this quantities using various methods can be found in the National Institute of Science and Technology database \cite{nist}. We have quoted the
result from the CCSD(T) method with daug-cc-pVTZ basis from this list in the above table. We find a good comparison between both the calculations. 
Similarly, a very precise experimental value of IP for Au$^-$ is reported \cite{hotop2} and we find that our result matches well with the experimental
value. We also compare our results with the other calculations that are reported using the DKnp Hamiltonian in the CCSD(T) method \cite{kello} and 
using the $j$-averaged relativistic pseudo-potential in the QCISD(T) method \cite{schwerdtfeger} in the above table. There are also another two more 
precise calculations of energies reported by Eliav {\it et al.} by considering four-component Dirac-Coulomb-Breit interaction Hamiltonian in the 
Fock-space RCCSD method with partial triples correction (RCCSD(T) method) \cite{eliav} and by Pasteka et al by using singles, doubles, triples, 
quadruples, and pentuples approximations in the relativistic equation-of-motion coupled-cluster method after including Breit and QED interactions 
with the Dirac-Coulomb Hamiltonian (DC-CCSDTQP$+$Breit$+$QED method) \cite{pasteka}. Our results are in agreement with the values reported in 
Refs. \cite{nist,pasteka,eliav}.   

\section{Summary}

We have employed relativistic many-body methods in the lower-order perturbation, random-phase approximation, and coupled-cluster theory frameworks 
by considering the four-component Dirac-Coulomb atomic Hamiltonian to analyze the trends in the electron correlation effects for the determination of
dipole polarizabilities of Cl$^-$ and Au$^-$ ions. The relativistic coupled-cluster theory is approximated at the singles and doubles excitations 
level. We have evaluated these values in the perturbative approach by preserving atomic spherical symmetry in contrast to the previous studies. 
We have compared our results with the previous calculations that were reported using the quasi-relativistic and scalar Douglas-Kroll spin-averaged 
(no-pair) Hamiltonians. We find reasonably good agreement among these results. We have also given contributions from various terms of the relativistic coupled-cluster theory and compared the trends
between both the considered negative ions. Moreover, we have analyzed ionization potentials of both the ions at different levels of approximations 
in the relativistic many-body methods, and compared with the available precise experimental results and calculations. Our results match well with 
the previous calculations suggesting that our methods are also reliable to produce these values. Our results can be further improved by including
higher-order relativistic effects and contributions from the full triple excitations through the relativistic coupled-cluster theory.

\section*{Acknowledgement}

 We acknowledge use of Vikram-100 HPC cluster of Physical Research Laboratory (PRL), Ahmedabad, India to carry out computations for this work.

\end{document}